\documentclass[12pt]{article}
\pdfoutput=1
\usepackage{wrapfig}
\usepackage [english]{babel}
\usepackage[applemac]{inputenc}
\usepackage[intlimits]{amsmath}
\usepackage{graphicx}
\usepackage[small]{caption}

\usepackage{setspace}
\allowdisplaybreaks
\DeclareGraphicsExtensions{.pdf,.png,.jpg}
\usepackage{xcolor}
\usepackage{float} 
\usepackage{slashed}
\usepackage{multirow}
\usepackage{setspace}
\usepackage{verbatim}
\usepackage{color}
\numberwithin{equation}{section}
\usepackage{hyperref}
\pdfpageattr {/Group << /S /Transparency /I true /CS /DeviceRGB>>}
\begin{document}

\title{Flavor corrections to the entanglement entropy}
\date{}

\author{Konstantina~Kontoudi$^{a,b}$ and Giuseppe~Policastro$^a$}

\maketitle

\thispagestyle{empty}

\begin{center}
$^a$ Laboratoire de Physique Th\'eorique, Ecole Normale Sup\'erieure,\\
 24 rue Lhomond, 75231 Paris Cedex 05,\\
 France (UMR du CNRS 8549)
~\\
\vspace{0.2cm}
$^b$ Laboratoire de Physique Th\'eorique et Hautes Energies,\\
 Universit\'e Pierre et Marie
Curie, 4 place Jussieu, \\
75252 Paris Cedex 05,
 France (UMR du CNRS 7589)
\end{center}

\vskip 1cm

\abstract{We consider the holographic entanglement entropy in  ${\cal N}=4$ SYM coupled to massive flavor degrees of freedom. The flavors are introduced by putting $D7$ branes in $AdS_5$. The resulting geometry including the backreaction of the branes is known in a perturbation expansion in the ratio $N_f/N_c$.  
We  consider the expansion to first order, and compute the entanglement entropy of a region of the boundary. We consider two different cases for the geometry of the region: a  slab and a ball. We find analytic solutions for the minimal surfaces in the bulk whose area gives the entropy, and analyze the structure of the UV divergence and the dependence on the masses. 
Our results confirm the general structure that was predicted by free field theory calculations, but with coefficients that depend on the coupling. }

\newpage

\tableofcontents

\setcounter{page}{1}
\setcounter{figure}{0}

\section{Introduction}

Entanglement is one of the most distinctive properties of quantum systems. Informally speaking, it corresponds to the fact that a measurement performed on a part of the system will affect another part, or alternatively it quantifies the amount of information on a subsystem that is accessible by performing measurements on another subsystem. 
There exist several measures of entanglement; the most commonly used is the   
{\emph {entanglement entropy}}. It can be naturally introduced in a quantum system divided into two subsystems A and B. Consider an observer that has only access to the subsystem A; the results of all the possible measurements he can make are encoded in the  reduced density matrix $\rho_{red}$ obtained by integrating out the degrees of freedom in B. The entanglement entropy (EE) of the subsystem A with B is defined as the von Neumann entropy associated to the reduced density matrix:
\begin{equation}
S_A=-\text{tr} (\rho_{red}\log\rho_{red}) \,.
\end{equation}
Very often one considers the case in which the subsystems are the degrees of freedom living  in different regions of space. 
The definition is completely general and can be in principle applied to any system, provided that the degrees of freedom are local, so that one can associate a Hilbert space to a given region of spacetime. 
On the other hand, EE is a very non-local observable, therefore it provides different information compared to local quantities such as correlators; for instance, it has been used as a probe of long-range topological order 
in two-dimensional systems with a mass gap \cite{Kitaev:2005dm}. It is also useful in many other contexts ranging from condensed matter to quantum information. 

EE has been the subject of intensive study in the last few years; its computation is generally a very challenging problem and few exact results are known. 
In a quantum field theory, EE is a UV divergent quantity and its computation requires the introduction of an ultraviolet regulator $a$. In terms of this cutoff the structure of the divergence, for a theory in $d+1$ spacetime dimensions, can be summarized as follows (see \cite{Calabrese:2004eu} for a more extended review of known properties):
\begin{equation}
\label{EEterms}
S_A=\frac{c_{d-1}}{a^{d-1}}+...+\frac{c_1}{a}+c_0\log a+S_f \,,
\end{equation}
where $S_f$ is  finite for $a \to 0$;  the coefficients $c_i$ depend in general on the geometric properties of the boundary surface $\Sigma$ separating the regions A and B, and have been computed in a limited number of cases (a review of the computational tools used to compute EE in free quantum field theories can be found in \cite{Casini:2009sr}).  
The leading divergent term  is proportional to the area of $\Sigma$, a fact known as the ``area law". 
Most of the terms in the expansion are actually ambiguous, as they are not invariant under a rescaling of the cutoff. One exception is the coefficient of $\log a$; in a conformal field theory, it has been shown to be related to the central charges appearing in the trace anomaly.

In a seminal paper \cite{Ryu:2006bv} Ryu and Takayanagi proposed a remarkably simple recipe for the computation of EE in theories with a holographic dual gravity description. The quantum field theory lives on the boundary of $AdS$;  consider a region of the boundary $A$ enclosed by the entangling surface $\partial A = \Sigma$.  According to the proposal, the EE of the region is proportional to the area $\mathcal{A}$ of a minimal surface that extends in the bulk of $AdS$ and whose restriction to the boundary of  $AdS$ is $\partial A$:
\begin{equation}
S=\frac{\mathcal A}{4G_N^{(d+2)}} \,.
\end{equation}
Among the various applications of this formula (see \cite{Nishioka:2009un}) it is worth mentioning the identification of the exact contribution of the central charges to the $\log a$  term \cite{Solodukhin:2008dh}.
This proposal has been proved in the case of a spherical entangling surface by mapping the problem of computing entanglement entropy to that of computing thermal entropy using a conformal transformation \cite{Casini:2011kv}. A more general proof, that should be applicable to any geometry of the entangling surface, has been recently proposed \cite{Lewkowycz:2013nqa} based on arguments about the solutions of gravitational theories with a boundary and their relation to the entropy of the density matrix.

The structure of the entanglement entropy presented in (\ref{EEterms}) is valid for conformal theories. 
When we move away from conformality the result can depend also on the intrinsic scales of the theory, such as masses. 
 We will concentrate on the corrections that appear in a massive deformation of a CFT.  
 Such corrections have been studied in \cite{Fursaev:2006ng}  for free scalar field theory with finite correlation length $\xi=1/m$ in a cubic region, and by \cite{Hertzberg:2010uv} in a waveguide geometry, i.e. a cylinder 
whose cross section has an arbitrary shape. It has been found that there is a finite contibution to the entropy of the form, in $d=3$, 
\begin{equation}
\label{logterm}
S_f=\frac{ A_\Sigma}{24 \pi} m^2 \log m +  f_0 \log m+f_1 m 
\end{equation}
where the coefficients $f_i$ depend on the geometrical characteristics of the waveguide, and $A_\Sigma$ is the area of the entangling surface. The terms appearing in (\ref{logterm}) are finite and independent of the ultraviolet regulator.  
They can be isolated from the UV-divergent part by taking derivatives with respect to the correlation length (see \cite{Liu:2012eea,Liu:2013una} for an alternative proposal for defining finite universal parts). 

In \cite{Hertzberg:2012mn} the first term of  (\ref{logterm}) has been computed perturbatively in a scalar field theory with $\phi^3$ and $\phi^4$ interactions, with the result that the structure remains the same but the bare mass is replaced by the renormalized mass. The same term has also been identified in a holographic computation of the entanglement entropy in \cite{Hung:2011ta} by introducing a massive scalar in AdS that sourced a relevant deformation of the CFT. Other computations have been done in string theory embedded backgrounds: the dual of $\mathcal{N}=2^*$ 
in \cite{Lewkowycz:2012qr}, and in the ABJM model with unquenched massive flavors in \cite{Bea:2013jxa}. We comment on their results in the conclusion section. 

In this paper we will consider another calculable example of EE in a massive field theory. We use the 
holographic prescription to compute the EE for $\mathcal N=4$ $U(N)$ SYM coupled to $N_f$ massive hypermultiplets; this is the theory that lives at the intersection of $N_c$ D3 and $N_f$ D7 branes \cite{Karch:2003nh}; in the regime 
$N_f \ll N_c$ the theory has a dual description in terms of probe D7 branes in $AdS_5$.  
In order to see the contribution of the flavor fields to EE we need to go beyond the probe (quenched) approximation and include the backreaction of the D7 branes (although at leading order it would also be possible to do the calculation remaining in the probe limit, see \cite{Chang:2013mca}). The backreacted solutions are known perturbatively in $\epsilon = N_f/N_c$ \cite{Bigazzi:2009bk}. 

We compute the EE in two cases, for an infinite region delimited by two hyperplanes (a slab) and for a ball, delimited by a sphere. 
We identify  the $\mu^2 \log \mu$ term and some of the power-law terms in (\ref{logterm}), thereby confirming Hertzberg's conjecture about the universality of these contributions. Moreover, given the consistent setup we use, we can compute the exact value of the coefficients; we found that they are modified from their free theory value. 

We should notice that even though we start from a consistent solution of supergravity, the dual theory is not in fact UV-complete: it has a Landau pole, as is reflected in the bad boundary behavior of the metric. This could be potentially problematic, and requires some special care when considering the boundary conditions and the counterterms. We found however that if the perturbative expansion is reorganized in terms of 
an effective coupling $\epsilon_q$ defined at the scale of the flavor fields' mass, the structure of the divergences is not different than what is expected in a renormalizable theory. 

The paper is organized as follows: in section \ref{sugra} we present the gravity solution dual to the D3/D7 system; in section \ref{holoEE} we start by reviewing the Ryu-Takayanagi prescription in the case of pure $AdS$, then we present our computation in the backreacted-branes geometry for the case of the slab and the ball; we conclude in section \ref{conc} by discussing our results, comparing them with previous results in the literature  and pointing out some possible extensions of our work.

\section{The backreacted D3/D7 geometry}\label{sugra}

We give a quick overview of the supergravity solution that we will use.  
The starting point is the $AdS^5\times S_5$ supergravity theory which is dual to  $\mathcal N=4$ SYM. Then we add flavors by introducing D7 branes and the backreaction of the branes is computed perturbatively in $\epsilon \sim N_f/N_c$ using the smearing technique (for an overview see \cite{Nunez:2010sf}). The branes are extended in space along the following directions:
\begin{center}
\begin{tabular}{c|cccccccccc}
 & $x_0$ &  $x_1$ & $x_2$ & $x_3$ & $x_4$ & $x_5$ & $x_6$ & $x_7$ & $x_8$ & $x_9$\\
 \hline
 D3 & $\times$ &  $\times$   & $\times$ &  $\times$  &  &   &  &   &  \\
 D7 & $\times$ &  $\times$   & $\times$ &  $\times$ & $\times$ &  $\times$   & $\times$ &  $\times$ & & 
\end{tabular}
\end{center}
Since the number of $D3$ branes is parametrically larger than the number of $D7$ branes, one can consider first the backraction of the $D3$ branes which results in the $AdS_5 \times S^5$ geometry. The $D7$ branes, considered as probes in the geometry, extend along the boundary directions of the $AdS_5$, along a part of the radial direction, and along an $S^3 \subset S^5$.  
The action of the coupled D3/D7 system in this regime is composed by the supergravity action in the $AdS$ background and the DBI action describing the flavor branes:
\begin{equation}
S=S_b+S_{fl}
\end{equation}
with:
\begin{align}
S_{b} =\frac{1}{2\kappa_{10}^2}\int  d^{10}x & \sqrt{-g_{10}} \left[R-\frac{1}{2}\partial_M\Phi\partial^{M}\Phi-\frac{1}{2}e^{2\Phi}F^2_{(1)}-\frac{1}{2}\frac{1}{5!}F^2_{(5)}\right]\\
S_{fl} & = -T_7\sum_{N_f}\left(\int d^8\ x e^{\Phi}\sqrt{-g_8}-\int C_8\right)\, .
\end{align}
If the D7 branes are localized in the directions transverse to their worldvolume, the equations of motion have  delta-function sources at the position of the branes and this makes them difficult to solve. The smearing technique consists in replacing the localized distribution of branes in the transverse space by a uniform brane density starting from a ``seed'' embedding and averaging using the symmetries of the internal space. 
In our case the D7 brane wraps an  $S^3 \subset S^5$. Even after averaging, there is a memory of the breaking of the isometries of the sphere that is reflected in a squashed sphere. This motivates the  
 following ansatz for the metric:
\begin{equation}
\label{metricAnsatz}
ds^2_{10}=h^{-1/2}(- dt^2+d\vec x^2_3)+h^{1/2}\left[F^2 d\rho^2+S^2ds^2_{CP^2}+F^2(d\tau+A_{CP^2})^2\right]
\end{equation}
\begin{gather*}
ds^2_{CP^2}=\frac{1}{4}d\chi^2+\frac{1}{4}\cos^2\frac{\chi}{2}(d\theta^2+\sin^2\theta d\varphi^2)+\frac{1}{4}\cos^2\frac{\chi}{2}\sin^2\frac{\chi}{2}(d\psi+\cos\theta d\varphi)^2\\
A_{CP^2}=\frac{1}{2}\cos^2\frac{\chi}{2}(d\psi+\cos\theta d\varphi)\\
\chi, \theta \in [0,\pi] \,, \phi,\theta \in [0,2\pi] \,, \psi \in [0,4\pi] 
\end{gather*}
The full solution contains also non-trivial RR forms but we will not mention them here since we will not need them. More details can be found in \cite{Bigazzi:2009bk} ; we report here the part of the results relevant for us.

All the fields depend only on the coordinate $\rho$ and we can find an one dimensional effective action by plugin in the ansatz in the action and integrating out the rest of the coordinates. The equations of motion arising from this action are equivalent to the following set of equations for a zero temperature setup:
\begin{equation}
\begin{split}
\quad \partial_{\rho} h=- \frac{Q_c}{S^4} ; & \quad \partial_{\rho}F = F\left(3-2\frac{F^2}{S^2}-\frac{Q_f}{2}e^{\Phi}\cos^4\frac{\chi}{2}\right)\\
\partial_{\rho}S=\frac{F^2}{S}; & \quad \partial_{\rho}\chi=-2 \tan\frac{\chi}{2}; \quad \partial_{\rho}\Phi=Q_f  e^{\Phi}\cos^4\frac{\chi}{2}
\end{split}
\end{equation}
where $\chi(\rho)$ is the ``seed'' brane embedding and the charges $Q_c$ and $Q_f$ are proportional to the number of colors and flavors respectively. \\
If the $D7$ branes are absent, the equations are solved by the $AdS$ metric. In the probe approximation, one sees that the branes extend along the radial direction from the boundary $\rho\rightarrow\infty$ to a finite  point $\rho_q$, related to the mass of the flavors in the boundary theory. This feature is preserved by the smearing procedure and persists after the backreaction. \\
The solution found for $\rho>\rho_q$ is:
\begin{gather*}
S_>  =\sqrt{\alpha'}e^{\rho}\left[1+\epsilon_*\left(\frac{1}{6}+\rho_*-\rho-\frac{1}{6}e^{6\rho_q-6\rho}-\frac{3}{2}e^{2\rho_q-2\rho}+\frac{3}{4}e^{4\rho_q-4\rho}-\frac{1}{4}e^{4\rho_q-4\rho_*}+e^{2\rho_q-2\rho_*}\right)\right]^{1/6}\\[0.1cm]
F_>=\sqrt{\alpha'}e^{\rho}\frac{\left[1+\epsilon_*\left(\rho_*-\rho-e^{2\rho_q-2\rho}+\frac{1}{4}e^{4\rho_q-4\rho}+e^{2\rho_q-2\rho_*}-\frac{1}{4}e^{4\rho_q-4\rho_*}\right)\right]^{1/2}}{\left[1+\epsilon_*\left(\frac{1}{6}+\rho_*-\rho-\frac{1}{6}e^{6\rho_q-6\rho}-\frac{3}{2}e^{2\rho_q-2\rho}+\frac{3}{4}e^{4\rho_q-4\rho}-\frac{1}{4}e^{4\rho_q-4\rho_*}+e^{2\rho_q-2\rho_*}\right)\right]^{1/3}}\\[0.1cm]
\Phi_>=\Phi_*-\log\left(1+\epsilon_*\left(\rho_*-\rho-e^{2\rho_q-2\rho}+\frac{1}{4}e^{4\rho_q-4\rho}+e^{2\rho_q-2\rho_*}-\frac{1}{4}e^{4\rho_q-4\rho_*}\right)\right)\, .
\end{gather*}
The dilaton diverges and the metric is not asymptotically $AdS$ when $\rho \to \infty$. The solution depends also on an arbitrary scale  $\rho_*$, an anchoring point at which the value of the dilaton is fixed; 
this point should also be viewed as the effective UV cutoff of the theory. Physically, this means that because of the Landau pole the theory can not be used for arbitrarily high energy. At the end of the calculation one should be able to send $\rho_* \to \infty$.

The solution in the region where the D7 branes do not extend i.e. for $\rho<\rho_q$, reads:
\begin{gather*}
\Phi_<=\Phi_q=\Phi_*-\log\left(1+\epsilon_*\left(\rho_*-\rho_q-\frac{3}{4}+e^{2\rho_q-2\rho_*}-\frac{1}{4}e^{4\rho_q-4\rho_*}\right)\right)\, ,\\[0.1cm]
S_<=F_<=\sqrt{\alpha'}e^{\rho}e^{-\frac{1}{6}(\Phi_q-\Phi_*)}\, .
\end{gather*}
For all values of the radial coordinate we can find $h$ by integrating the equation
\begin{equation}
\label{heq}
\frac{dh}{d\rho}=-\frac{Q_c}{S^4}
\end{equation}
with $Q_c$ being proportional to the number of colors $N_c$. The perturbation parameter is given by:
\begin{equation}
\epsilon_*=\frac{1}{8\pi^2}\lambda_*\frac{N_f}{N_c}
\end{equation}
where $\lambda_*$ is the 't Hooft coupling at the $\rho_*$ scale. For our purposes though it is preferable to express the solution in terms of a perturbation parameter fixed at the flavor mass scale given by:
\begin{equation}
\epsilon_q=\epsilon_*e^{\Phi_q-\Phi_*}\, .
\end{equation}
Since we are interested in computing quantities at the scale lower than the mass of the flavors, $\epsilon_q$ is the effective expansion parameter that has to be kept small; the residual dependence on the cutoff scale leads to subleading contributions that can be suppressed sending $\rho_* \to \infty$.  This observation was done in \cite{Magana:2012kh} in considering the dynamics of probe quarks in the unquenched flavored plasma; we verified explicitly that the same happens in our case. 

Fixing the reparametrization invariance of the metric we can define a new coordinate $z$ by imposing that $h$ takes the form:
\begin{equation}
\label{cond}
h(z)=\frac{z^4}{R^4};\quad \quad R^4\equiv \frac{1}{4}Q_c\, .
\end{equation}
This form is the same as in the unflavored case and it is convenient for comparing our results with the pure $AdS$ case. Imposing this condition and integrating equation (\ref{heq}) order by order we find an expression for $z(\rho)$. We fix the additive integration constant in $h$ by requiring that $z\rightarrow 0$ when $\rho\rightarrow \infty$. Then we have for $\rho>\rho_q$:
\begin{equation}
\begin{split}
\label{zrhobig}
z_>(\rho)= & \frac{e^{-\rho}R^2}{\sqrt {\alpha'}}\left[1+\frac{\epsilon_q}{720} \left(\frac{8 e^{-6 \rho } R^{12}}{\alpha ^{\prime\ 3} z_q^6}-\frac{45 e^{-4 \rho }
   R^8}{\alpha ^{\prime 2} z_q^4}+\frac{30 e^{-4 \rho _*} R^8}{\alpha^{ \prime 2}
   z_q^4}\right.\right.\\
& \left.\left.+\frac{120 e^{-2 \rho } R^4}{\alpha ' z_q^2}-\frac{120 e^{-2 \rho _*} R^4}{\alpha '
   z_q^2}+120 \rho -120 \rho _*+10\right)\right]
\end{split}
\end{equation}
where we defined $z_q=z(\rho_q)$. Now we can invert this relation to obtain $F_>(z)$ and $S_>(z)$ as expansions up to first order in $\epsilon_q$:
\begin{equation*}
\begin{split}
F_>(z)= &\frac{R^2}{z}+ \frac{R^2 \epsilon_q}{240 z
   z_q^6} \left(-45 z^4 z_q^2+40 z^2 z_q^4-10 z_q^6+16 z^6\right)
\end{split}
\end{equation*}
\begin{equation*}
\begin{split}
S_>(z)= &\frac{R^2}{z}+\frac{R^2 \epsilon_q}{240 z
   z_q^6} \left(15 z^4 z_q^2-20 z^2 z_q^4+10 z_q^6-4 z^6\right)\, .
\end{split}
\end{equation*}
Imposing continuity of the function $h$ at $\rho=\rho_q$ we obtain the following expressions for the coordinate $z$ and for the functions $F_<(z)$ and $S_<(z)$ for $\rho<\rho_q$:
\begin{equation}
\begin{split}
\label{zrhosmall}
z_<(\rho)=  &\frac{e^{-\rho}R^2}{\sqrt {\alpha'}}\left[1+\epsilon_q\left(\frac{e^{-4 \rho _*} R^8}{24 \alpha ^2 z_q^4}+\frac{\alpha ^2 e^{4 \rho } z_q^4}{240 R^8}-\frac{e^{-2 \rho _*} R^4}{6 \alpha  z_q^2}-\frac{1}{6} \log
   \left(\frac{\sqrt{\alpha' } z_q}{R^2}\right)-\frac{\rho _*}{6}+\frac{1}{8}\right)\right]\, ,
\end{split}
\end{equation}
\begin{equation*}
\begin{split}
F_<(z)=S_<(z)= & \frac{R^2}{z}+\epsilon_q\frac{R^2 z_q^4}{720 z^5 } \, .
\end{split}
\end{equation*}
This completes the discussion of the ingredients necessary for the computation of the entanglement entropy.

\section{Holographic entanglement entropy computation}\label{holoEE}
\subsection{Review of the pure AdS case}
We recall here the computation of the entanglement entropy for a slab and a ball geometry in pure $AdS_{d+2}$. The metric is given by:
\begin{equation}
ds^2=\frac{R^2}{z^2}\left(-dt^2+\sum_{i=1}^{d}dx_i^2+dz\right)\, .
\end{equation}
The slab is defined on a constant time slice on the boundary as:
\begin{equation*}
x_1 \in [-l/2,l/2] \ ; \quad x_{2,3,...,d} \in (-\infty,\infty)
\end{equation*}
\begin{figure}[H] 
 \center
  \includegraphics[width=10cm,viewport= 70 300 350 520]{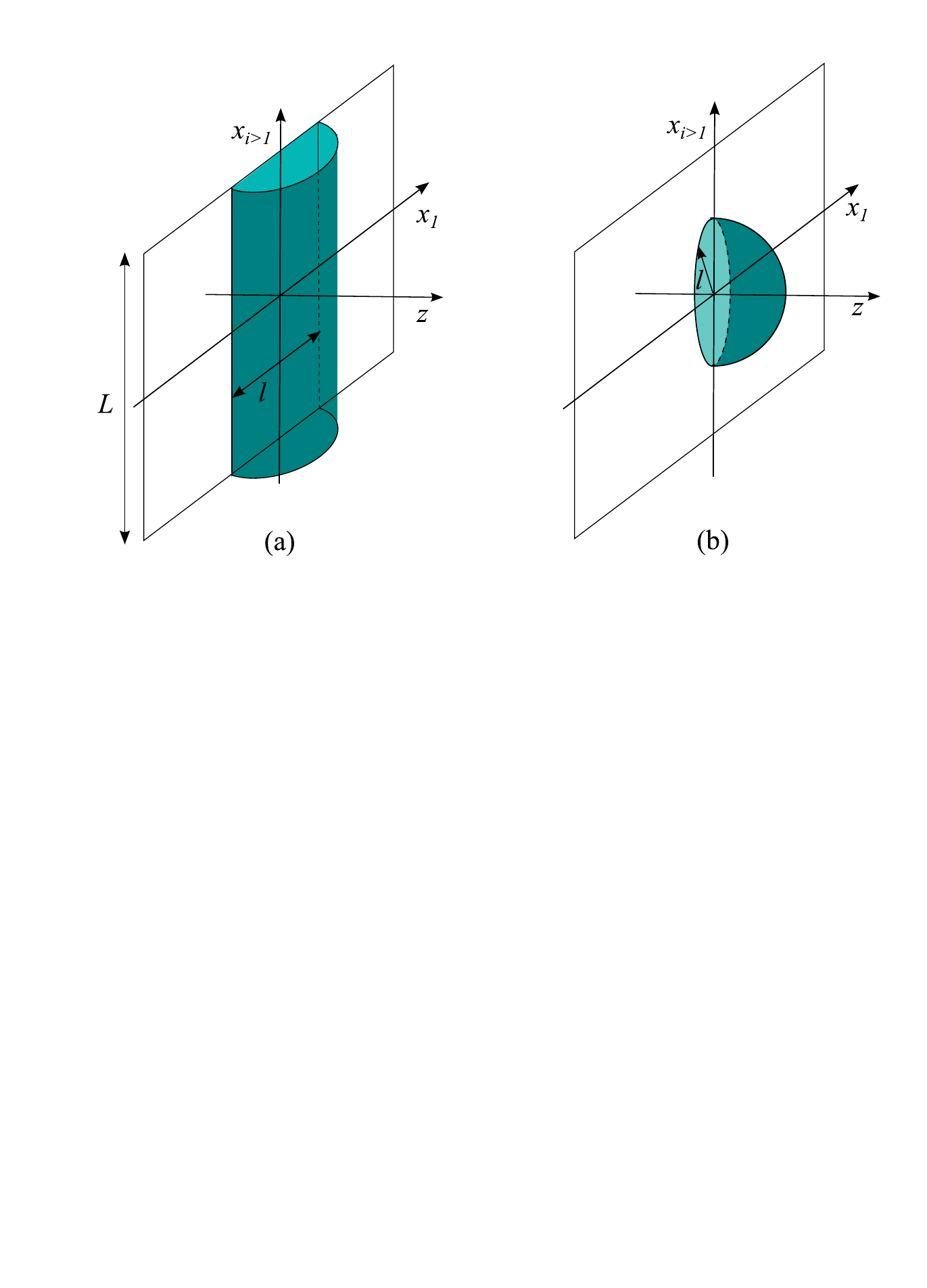}
  \caption{The slab geometry (a) and the ball geometry (b) and the corresponding minimal surfaces in $AdS$ space.}
  \label{slab}
 \end{figure}
We will use the regularized length L for the infinite directions as shown in the geometrical construction in fig. \ref{slab} (a). The holographic entanglement entropy can be computed as the area $\mathcal A$ of the minimal surface extending in the $AdS$ bulk and whose boundary lies on the entangling surface separating the slab and the rest of the boundary.
We start by minimizing the area functional for the surface extending in the bulk. Choosing an embedding of the form $z=z(x_1)=z(x)$ for the surface we have:
\begin{equation}
S_{\text{area}}=R^dL^{d-1}\int_{-\ell/2}^{\ell/2} dx\ \frac{\sqrt{1+z'^2}}{z^d}\, .
\end{equation}
Given that the integrand does not depend explicitly on $x$ we can compute the constant of motion and get
 \begin{equation}
 \frac{dz}{dx}=\frac{\sqrt{\tilde z^{2d}-z^{2d}}}{z^d}
 \end{equation}
where $\tilde z$ is the turning point of the surface. The minimal area is therefore given by:
\begin{equation}
\mathcal A=2 R^dL^{d-1}\int dz\ \frac{\tilde z^d}{z^d\sqrt{\tilde z^{2d}-z^{2d}}}\, .
\end{equation}
To compute the integral we need to introduce a UV cutoff $a$ and also satisfy the constraint:
\begin{equation}
\frac{\ell}{2}=\int_{-\ell/2}^0 dx=R^2\frac{\sqrt{\pi}\ \Gamma\left(\frac{d+1}{2d}\right)}{\Gamma\left(\frac{1}{2d}\right)}\tilde z\, .
\end{equation}
The area of the minimal surface after regularization  is given by:
\begin{equation}
\label{EEAdS}
\mathcal A_{AdS}= \frac{2R^d}{d-1}\left(\frac{L}{a}\right)^{d-1}-\frac{2^d\pi^{d/2}R^d}{d-1}\left(\frac{\Gamma\left(\frac{1+d}{2d}\right)}{\Gamma\left(\frac{1}{2d}\right)}\right)^d\left(\frac{L}{\ell}\right)^{d-1}\, .
\end{equation}

We move on now to the computation for the ball geometry where the entangling surface is a sphere of radius $\ell$ (fig. \ref{slab} (b)). It is convenient to write the metric in spherical coordinates; introducing the coordinate $r^2=\sum_{i=1}^dx_i$ the metric becomes
\begin{equation}
ds^2=\frac{R^2}{z^2}\left(-dt^2+dr^2+r^2d\Omega^2_{d-1}+dz^2\right)\, .
\end{equation}
Choosing an embedding of the form $r=r(z)$ the area functional is given by
\begin{equation}
S_{\text{area}}=R^d\text{vol}(S^{d-1})\int dz\ \frac{r^{d-1}}{z^d}\sqrt{1+r'^2}\, .
\end{equation}
The equations of motion of this area functional admit the solution
\begin{equation}
r^2+z^2=\ell^2\, .
\end{equation}
The minimal area is therefore given by:
\begin{equation}
\mathcal A=R^d\text{vol}(S^{d-1})\int_{a/\ell}^1du\ \frac{(1-u^2)^{\frac{d-2}{2}}}{u^d}
\end{equation}
where a is the UV cutoff. For small values of the cutoff and for d odd the minimal area can be expressed as a series of the following form:
\begin{equation}
\mathcal A=\frac{2\pi^{d/2}R^d}{\Gamma(d/2)}\left[p_1\left(\frac{\ell}{a}\right)^{d-1}+p_3\left(\frac{\ell}{a}\right)^{d-3}+...+p_{d-2}\left(\frac{\ell}{a}\right)^2+p_o\log\frac{\ell}{a}\right]\, .
\end{equation}
The values of the coefficients for $d=3$, which will be of interest to us are $p_1=1/2$ and $p_o=-1/2$.

\subsection{Flavor corrections : the slab}

We move now to the computation of the entanglement in the backreacted D3/D7 geometry given in 
(\ref{metricAnsatz}). For the case of the slab geometry, we choose an embedding of the form $\rho=\rho(x)$; the area functional of the surface is then given by\footnote{We divide the area by $R^5 \text{vol}(S^5)$ to make the results comparable with the AdS case where there is no internal five sphere.}:
\begin{equation}
 S_{\text{area}}=\frac{L^2}{R^5} \int_{-\ell/2}^{\ell/2} dx\ h^{1/2}FS^4\sqrt{1+hF^2\rho^{\prime 2}}\, .
\end{equation}
The embedding function satisfies the equation
\begin{equation}
 \frac{d\rho}{dx}= -\frac{\sqrt{hF^2S^8-\tilde h\tilde F^2\tilde S^8}}{\sqrt{h\tilde h}F\tilde F\tilde S^4}
\end{equation}
where we denote $\tilde \rho$ the turning point of the surface and the tilded functions are the values of the functions at the turning point. Using this relation the minimal area can be computed as follows:
\begin{gather}
\mathcal A=2\frac{L^2}{R^5}\int_{\tilde\rho}^{\infty}d\rho\ \frac{h^{3/2}F^3S^8}{\sqrt{hF^2S^8-\tilde h\tilde F^2\tilde S^8}}\, ,\\
\ell=2\int_{\tilde\rho}^{\infty}d\rho\ \frac{\sqrt{h\tilde h}F\tilde F\tilde S^4}{\sqrt{hF^2S^8-\tilde h\tilde F^2\tilde S^8}}\, .
\end{gather}
For convenience we switch to the $z$ coordinates given in terms of $\rho$ by eq. (\ref{zrhobig}) and (\ref{zrhosmall}) in the regions $\rho>\rho_q$ and $\rho<\rho_q$ respectively. To regularize the area integral we introduce a UV cutoff at $z=a$. We compute the width of the slab and the minimal area to first order in the perturbation parameter $\epsilon_q$:
\begin{gather}
 \ell=\ell_0+\epsilon_q\ell_1\\
\mathcal A=\mathcal A_0+\epsilon_q \mathcal A_1\, .
\end{gather}

\subsubsection*{I. Turning point located at $\tilde\rho>\rho_q$ ($\tilde z < z_q$)}
We start by computing the entropy for the case $\tilde\rho>\rho_q$ i.e. the turning point is located in the region where the D7 branes extend. We can express both the length $\ell$ and the area ${\mathcal A}$ in terms of the parameter $b=\tilde z/z_q$.
\begin{equation}
\label{lofc}
 \frac{\ell}{z_q}= \gamma_1 b  + \epsilon_q\left[\frac{1}{720} b^5 \left(48 \gamma _3-15 \gamma _2\right)+\frac{1}{720} b^3 \left(40 \gamma _2-160\right)+\frac{b \gamma _2}{8}\right]\, ,
\end{equation}
\begin{equation}
\begin{split}
\frac{z_q^2}{L^2R^3}\mathcal A^I= & -\frac{\gamma_1}{2b^2}+ \epsilon_q\left\{\frac{1}{144} b^2 \left(24 \gamma _3-3 \gamma _2\right)-\frac{\gamma _2}{8 b^2}+\frac{1}{144} \left[8 \gamma _2-16 (6 \log (b z_q)+1+\log
   4)\right]\right\}
\end{split}
\end{equation}
where
\begin{equation*}
\gamma_1=\frac{2\sqrt{\pi}\Gamma(2/3)}{\Gamma(1/6)} ;\quad\quad \gamma_2=\frac{ \Gamma(2/3) \Gamma(5/6)}{\sqrt{\pi}}  ;\quad\quad  \gamma_3=\frac{\Gamma\left(1/3\right)\Gamma\left(7/6\right)}{\sqrt{\pi}}\, .
\end{equation*}
The divergent piece of the area is given by:
\begin{equation}
 \mathcal A_{\text{div}}=\frac{L^2R^3}{a^2}-\epsilon_qL^2R^3\left[\frac{1}{4a^2}-\frac{2}{3z_q^2}\log a\right]\, .
\end{equation}
The zeroth order term of the area matches the result for the $AdS$ case eq. (\ref{EEAdS}) for $d=3$ as expected. 
To express the area in terms of $\ell$ we can pertubatively invert the relation (\ref{lofc}) which leads to:\\
\begin{equation}
\begin{split}
\mathcal A^I(\ell)= &-\frac{\gamma_1^3 L^2 R^3}{2 \ell^2}+\epsilon_q L^2 R^3\left[\frac{6 \log\gamma _2+1-2\log2}{9 z_q^2}+\frac{\gamma _3 \ell ^2}{10 \gamma _2^2 z_q^4}-\frac{\gamma _2^3}{4 \ell ^2}\right]\, .
\end{split}
\end{equation}

\subsubsection*{II. Turning point located at $\tilde \rho<\rho_q$ ($\tilde z >  z_q$)}

To compute the length and area integrals in this case we must split them in two parts: one from the boundary to $\rho_q$ and another one from $\rho_q$ to the turning point.
The results that we find for the length and area are:  
\begin{equation}
\begin{split}
\frac{\ell}{z_q}& =   \gamma_1 b  + \frac{\epsilon_q}{2160}\left[-\frac{30}{b^7}-\frac{15 \gamma _1}{b^3}-480 b^3+\frac{30}{b^3}+48 b^5 B\left({\frac{1}{b^6}};\frac{1}{3},\frac{1}{2}\right)\right.\\
   &\left.+\frac{6 \sqrt{b^2-1}
   \left(56 b^4+71 b^2+31\right)}{\sqrt{b^4+b^2+1}}+\left(-15 b^5+40 b^3+\frac{5}{b^3}+90 b\right)
   B\left({\frac{1}{b^6}};\frac{2}{3},\frac{1}{2}\right)\right]\, ,
\end{split}
\end{equation}\\
\begin{equation}
\begin{split}
\frac{z_q^2}{L^2R^3}\mathcal A^{II}= -\frac{\gamma_1}{2b^2}+ & \frac{\epsilon_q}{432 b^7}\left[24 b^9 B\left({\frac{1}{b^6}};\frac{1}{3},\frac{1}{2}\right)+\left(-3 b^8+8 b^6-18 b^4+1\right) b
   B\left({\frac{1}{b^6}};\frac{2}{3},\frac{1}{2}\right)\right.\\
   &\left.-48 b^7-48 b^7 \left(-2 \cosh ^{-1}\left(b^3\right)+6 \log (b z_q)+2\log
   2\right)\right.\\
   &\left.-\frac{6\sqrt{b^2-1} \left(12 b^8+9 b^6+17 b^4-b^2-1\right)}{\sqrt{b^4+b^2+1}}-3 b \gamma _1\right]\, .
 \end{split}
\end{equation}
where $B(z;a,b)$ is the incomplete Beta function defined as:
\begin{equation}
B(z;a,b)=\int_0^z t^{a-1}(1-t)^{b-1}dt\, .
\end{equation}
The counterterms used for the regularization of the area are the same as for the surface extending only in the $\rho>\rho_q$ region since the fact that the surface extends further in the interior does not affect the ultraviolet behavior of the integrals. 
%
Now we can invert again the relation $\ell(b)$ to express the area in terms of $\ell$:
\begin{multline}\label{areaslab}
\frac{z_q^2}{L^2R^3}\mathcal A^{II}(\ell)  = -\frac{\gamma_1}{2b^2}+\frac{\epsilon_q}{720} \left[-\frac{90 \gamma _1^6 z_q^6}{\ell ^6} \, _2F_1\left(\frac{1}{2},\frac{2}{3};\frac{5}{3};\frac{z_q^6 \gamma _1^6}{\ell
   ^6}\right)+72 \, _2F_1\left(\frac{1}{3},\frac{1}{2};\frac{4}{3};\frac{z_q^6 \gamma _1^6}{\ell ^6}\right)\right.\\
    +\frac{10 \gamma
   _1^{10} z_q^{10}}{\ell ^{10}}-\frac{10 \gamma _1^6 z_q^6}{\ell ^6}+160 \cosh ^{-1}\left(\frac{\ell ^3}{\gamma _1^3
   z_q^3}\right)+\frac{10 \gamma _1^4 z_q^4 \sqrt{\ell ^6-\gamma _1^6 z_q^6}}{\ell ^7}+80 \log \left(\frac{\gamma _1^6}{4 \ell ^6}\right)\\
 \left.   -\frac{232 \gamma _1^2 z_q^2 \sqrt{\ell ^6-\gamma
   _1^6 z_q^6}}{\ell  \left(\gamma _1^4 z_q^4+\gamma _1^2 \ell ^2 z_q^2+\ell ^4\right)}-\frac{232 \ell  \sqrt{\ell ^6-\gamma _1^6
   z_q^6}}{\gamma _1^4 z_q^4+\gamma _1^2 \ell ^2 z_q^2+\ell ^4}-\frac{242 \gamma _1^4 z_q^4 \sqrt{\ell ^6-\gamma _1^6 z_q^6}}{\ell ^3
   \left(\gamma _1^4 z_q^4+\gamma _1^2 \ell ^2 z_q^2+\ell ^4\right)}+80\right]\, .
\end{multline}
We are interested in the behavior of the theory for large values of $\ell$ in order to probe the cutoff independent mass corrections to the entanglement entropy. Following Hertzberg and Wilczek \cite{Hertzberg:2010uv}, we can extract these cutoff-independent contributions; identifying $\xi^{-1} = m = 1/z_q$, the cutoff independent part is 
 \begin{equation}\label{Sxidef}
 S_\xi = (-\xi^{-2})^2 \frac{\partial S}{\partial (\xi^{-2})^2} \,.
 \end{equation}
 We can check that indeed this quantity is UV-finite, and it is a function of $\Lambda^2 \equiv \ell^2/z_q^2$. 
 The large $\Lambda$ expansion, $\ell\gg\xi$, reveals the following term:
\begin{equation}
\label{Sxislab}
S_{\xi}\approx\epsilon_q\frac{L^2R^3}{G_N}\frac{1}{3\xi^2}=\frac{1}{2 \pi^2}\lambda_qN_fN_c\frac{A_\Sigma}{48\pi\xi^2}\, .
\end{equation}
Note that an entropy of the form 
\begin{equation}
S=-\frac{A_\Sigma}{24\pi}\frac{1}{\xi^2}\log\xi-\frac{4b_1}{\xi}+2b_0\log\xi
\end{equation}
produces an $S_\xi$ of the following form
\begin{equation}
\label{Sxi}
S_\xi =  \frac{{\cal A}_\Sigma}{48\pi \xi^2} + \frac{b_1}{\xi} + b_0 \, .
\end{equation}
Therefore the term that we found for the slab geometry corresponds to the $m^2 \log m$ term in (\ref{logterm}).  The constant term and the $1/\xi$ terms are missing compared to (\ref{Sxi}) which was identified as the free field theory result in a waveguide geometry in \cite{Hertzberg:2010uv}. The coefficient $b_1$ is related to the perimeter of the waveguide and the $b_0$ is related to curvature; the fact that there are no analogs of these geometric quantities in the slab geometry is probably the reason of the absence of these terms.

\subsection{Flavor corrections : the ball} 

In this section we consider the case where the entangling surface is a sphere of radius $\ell$. 
The embedding of the minimal surface is given in terms of a function $r(z)$ where $r$ is the radial coordinate in the boundary, $r^2 = \sum x_i^2$. It is convenient to make a change of variable to 
$r^2 = y(z)^2 - z^2$. The AdS solution then reads simply $y = \textrm{const} = \ell$.  
The area functional is now 
\begin{equation} 
\label{areaSph}
S_{\text{area}} = \frac{4 \pi}{R^5}\int dz\  h^{1/2} S^4 F (y^2-z^2)^{1/2} \sqrt{(y y' - z)^2+ h F^2 \rho'(z)^2 (y^2 - z^2)}\, .
\end{equation}
The corresponding equations of motion
\begin{equation}
\begin{split}
& 4 F^5 h^3 S \rho ^{\prime\, 4} \left(y^2-z^2\right)+2 F^3 h^2 \rho ' \left\{\rho ' S \left[z^2 \left(y^{\prime\, 2}+2\right)-y^3y''+y z \left(z y''-6 y'\right)+y^2 \left(2 y^{\prime\, 2}+1\right)\right] \right.\\
&\left. -(4 S' \rho'-S \rho '')\left(z^2-y^2\right) \left(z-y y'\right)\right\}+2 h \left(S F'+4 F S'\right) \left(z-y y'\right)^3+F S h' \left(z-y y'\right)^3=0
\end{split}   
\end{equation}
can be solved at first order in $\epsilon_q$. We denote the perturbed solution by 
$y = y_0 + \epsilon_q y_1$. 
Again we have to distinguish the case where the surface extends in the bulk only in the region $\rho>\rho_q$ 
from the case where it goes further in the bulk. 

\subsubsection*{I. Turning point located at $\tilde\rho>\rho_q$ ($\tilde z < z_q$)}

In this case the perturbed solution is 
\begin{eqnarray} \label{y1sollar}
y_1^{I}(z)&  = & w(z) + \frac{C_1^I \left(z^2-2 \ell^2\right)}{\sqrt{\ell^2-z^2}} + C_2^I \, , \\ 
w(z) &\equiv &   \frac{4 \ell^3 \log z}{3 z_q^2}+\frac{\left(4 \ell^4-2 \ell^2 z^2\right)
   \log \left(\frac{\sqrt{\ell^2-z^2}+\ell}{z}\right)}{3 z_q^2 \sqrt{\ell^2-z^2}}-\frac{z^6}{80 \ell
   z_q^4}-\frac{\ell z^4}{30 z_q^4}+\frac{z^4}{18 \ell z_q^2}+\frac{\ell z^2}{3
   z_q^2}-\frac{z^2}{8 \ell} \,. \nonumber
\end{eqnarray}
 
In order for the solution to be regular at $z=\ell$ we must set  $C_1^I=0$. The other constant is fixed by the boundary condition  $y_1^I(z=0) =0$: 
 \begin{equation*}
 C_2^I = - \frac{4 \ell^3}{3 z_q^2} \log (2 \ell) \,.
 \end{equation*}
 The integral for the area can be calculated analytically, and we have 
 \begin{equation}\label{spherediv}
 {\cal A}^I = 4 \pi R^3 \left[ \frac12 \frac{\ell^2}{a^2} - \frac12 \log (\frac{\ell}{a}) + \epsilon_q \left( \frac{\ell^2}{8 a^2} + \frac{4 \ell^2 + 3 z_q^2}{12 z_q^2} \log(\frac{a}{2 \ell}) + \frac{\ell^4}{30 z_q^4} + \frac{7 \ell^2}{18 z_q^2} - \frac{1}{16}
 \right)\right]\, .
 \end{equation}
 
 \subsubsection*{II. Turning point located at $\tilde\rho < \rho_q$ ($\tilde z > z_q$)}
 
 In this case the embedding is described by two different functions: 
 
 \begin{equation} \label{y1sollar2} 
y_1^{II} = \left\{ 
\begin{array}{ll} 
w(z) + C_1^{II} \displaystyle{ \frac{z^2-2 \ell^2}{\sqrt{\ell^2-z^2}}} + C_2^{II} \, , & \, \, 0 < z  < z_q \\ [0.6cm]
\displaystyle{\frac{z_q^4}{144 \ell z^2}} + D_1 \left( \sqrt{z^2-\ell^2}-\displaystyle{\frac{\ell^2 }{\sqrt{z^2-\ell^2}}} \right)+D_2 \,,  & \, \, z_q < z < \tilde z 
\end{array} \right. 
\end{equation} 
As in case I, regularity of the solution at $z=\ell$ fixes $D_1=0$, and the boundary condition $y_1^{II}(z=0) = 0$ fixes 
\begin{equation*}
C_1^{II}= \frac{C_2^{II}}{2 \ell }+\frac{2 \ell ^2 \log (2 \ell )}{3 z_q^2}\, .
\end{equation*}

 The two remaining constants are fixed by requiring the continuity of the solution and of the first derivative at the matching point $z=z_q$. Notice that even though we are matching the solutions in two different regions, 
 the point $z_q$ is not really a boundary as the metric is smooth across this point, therefore there is no ``refraction" and the geodesics are smooth curves. The matching condition gives:
 \begin{equation}
 \begin{split}
 C_2^{II} & = \displaystyle{\frac{4}{45 z_q^4} \left(\sqrt{\ell ^2-z_q^2} \left(14 \ell ^2 z_q^2-2 z_q^4+3 \ell ^4\right)+15 \ell ^3 z_q^2 \log \left(\frac{2 \ell 
   z_q}{\sqrt{\ell ^2-z_q^2}+\ell }\right)\right)}\,, \\
 D_2 & =\frac{1}{90 z_q^2 \left(\ell  \left(\sqrt{\ell ^2-z_q^2}+\ell \right)-z_q^2\right)}\Bigg[-83 \ell ^2 z_q^2 \sqrt{\ell ^2-z_q^2}+16 z_q^4 \sqrt{\ell ^2-z_q^2}\\
 & + 12 \ell ^4 (10\log2-1) \sqrt{\ell
   ^2-z_q^2}-120 \ell ^3 \left(\ell  \left(\sqrt{\ell ^2-z_q^2}+\ell \right)-z_q^2\right) \log \left(\sqrt{1-\frac{z_q^2}{\ell ^2}}+1\right)\\
&  -29 \ell 
   z_q^4+12 \ell ^5 (1+10\log2)+\ell ^3 (17-120 \log 2) z_q^2\Bigg].
    \end{split}
 \end{equation}
Once again the integration can be performed analytically, with the following result
\begin{equation}\label{areaball}
\begin{split}
\frac{{\cal A}^{II}}{ 4 \pi R^3} = & \frac{1}{2}\frac{\ell^2}{a^2}-\frac{1}{2}\log\frac{\ell}{a}  +\epsilon_q\Bigg\{
\frac{\ell^2}{8 a^2} 
-   \frac{1}{720 \ell  z_q^4}\Bigg[4 \ell ^2 z_q^2 \Bigg(60 \ell  \log\frac{a \left(\sqrt{\ell ^2-z_q^2}+\ell \right)}{2 \ell  z_q}\\
& -83 \sqrt{\ell ^2-z_q^2}+70 \ell
   \Bigg)+z_q^4 \Bigg(180 \ell \log\frac{a \left(\sqrt{\ell ^2-z_q^2}+\ell \right)}{2 \ell  z_q}-45\ell-64 \sqrt{\ell
   ^2-z_q^2}\Bigg)\\
   & +24 \ell ^4 \left(\ell -\sqrt{\ell ^2-z_q^2}\right)\Bigg]\Bigg\}\, .
   \end{split}
\end{equation}  
The turning point, both in case I and II, is modified from its zeroth order value and is determined by $\tilde z = \ell + \epsilon_q y_1 (\ell)$. However this shift does not affect the area, to first order in $\epsilon_q$, 
since the integrand of the action functional evaluated on the zeroth order solution vanishes at $\tilde z$. 

The divergent terms in the last formula are the same as in (\ref{spherediv}), as it must be since the divergence comes only from the $z\sim0$ region. We extract the mass-dependent universal part using (\ref{Sxidef}). Again we find that it is a function of $\Lambda^2 \equiv \ell^2/z_q^2$. 
In the limit of large $\Lambda$, $\ell \gg \xi$, it has an expansion
\begin{equation}\label{Sxiball}
S_\xi \approx 4 \pi R^3 \epsilon_q \, (\frac16 \Lambda^2 - \frac18) \, = \frac{\lambda_q}{2 \pi^2} N_f N_c \left(\frac{{\cal A}_\Sigma}{48 \pi \xi^2} - \frac{1}{16} \right) \,.
\end{equation}
Comparing with (\ref{Sxi}), we see that we find the leading term and the constant term, while the 
term proportional to $1/\xi$ is once again missing.

\section{Conclusions}\label{conc}

We have computed the corrections to the entanglement entropy due to the massive flavor fields coupled to ${\cal N}=4$ SYM in 3+1 dimensions; from these we could extract the UV-divergent terms and the universal mass-dependent finite terms. The main results of this paper are contained in 
eqs. (\ref{areaslab}),(\ref{Sxislab}),(\ref{areaball}),(\ref{Sxiball}), giving the exact result for the area and the finite mass-dependent terms for the slab and the ball, respectively. 

It is instructive to compare what we found with the previously known results. As already mentioned, the mass-dependent terms have been computed for the first time 
in \cite{Hertzberg:2010uv} for a free field; the contribution is 
$$S_{free} \sim \gamma \, {\cal A}_\Sigma \, m^2 \log m$$ 
with $\gamma = \displaystyle{\frac{1}{24\pi}}$ for a scalar, and $\gamma = \displaystyle{\frac{1}{48\pi}}$ for a Dirac fermion (in 3+1 dimensions). 

Subsequently, in \cite{Hertzberg:2012mn} the coupling constant dependence of the coefficient $\gamma$ was studied at one loop in perturbation theory for cubic and quartic scalar interactions; the result was that  $\gamma$ is unchanged if $m$ is taken to be the renormalized mass. 

In \cite{Lewkowycz:2012qr} the entanglement was computed in the ${\cal N}=2^*$ SYM theory, which is a deformation of ${\cal N}=4$ SYM by relevant operators $m_b^2 {\cal O}_2 + m_f {\cal O}_3$ that give mass to the scalars and  to the fermions. The theory is supersymmetric only for $m_b=m_f$, otherwise susy is broken and for $m_b > m_f$ there is a tachyonic mode, however the computation of the entanglement is insensitive to these issues.  
The result they found is that adding the operator $m_f {\cal O}_3$, 
$${\cal O}_3 = -i \textrm{Tr} \,  \psi_1 \psi_2 + \frac{2}{3} m_f \sum_{i=1}^3 \textrm{Tr} | \phi_i |^2  \,,$$ 
which gives mass $m_f$ to fermions and $2/3 m_f^2$ to bosons, the entanglement computed holographically is 
$$S_{{\cal N}=2^*} \sim \frac{N^2}{12\pi} {\cal A}_\Sigma \, m^2 \log m \,.$$
It can be easily verified that the computation at weak coupling would give instead a factor of $1/4 \pi$. There is then a disagreement between weak and strong coupling, the two results differ by a finite multiplicative factor. 

In the theory we considered, the massive degrees of freedom are ${\cal N}=2$ hypermultiplets $Q_I, \tilde Q^I$ in the bifundamental representation of $U(N_f) \times U(N_c)$. Each hypermultiplet contains two complex scalars and two Weyl fermions. The weak-coupling computation would give then 
$$S_{{\cal N}=2} \sim \frac{6 N_f N_c}{24 \pi} {\cal A}_\Sigma \, m^2 \log m \,.$$
Comparing with (\ref{Sxislab}) or (\ref{Sxiball}) we see that like for the ${\cal N}=2^*$ case we have a disagreement: at strong coupling the factor 6  in the numerator is replaced by $\lambda_q / 2\pi$. 
These results cast some doubt on the conjecture of \cite{Hertzberg:2012mn} even though both cases are not very conclusive: in \cite{Lewkowycz:2012qr} the operator ${\cal O}_3$ actually does not contain only mass terms but also Yukawa couplings (that we didn't write). In our case also one source of ambiguity comes from the difficulty in defining precisely the flavor mass, since the quarks are not gauge-invariant operators 
and one should more properly talk about meson masses. 

It would be interesting nevertheless to pursue the perturbative computation of \cite{Hertzberg:2012mn} to higher order, to see if the discrepancy persists. 

Another piece of evidence comes from the recent work \cite{Bea:2013jxa} in which they considered the three-dimensional ABJM Chern-Simons matter
theory with unquenched massive flavors. The flavor degrees of freedom are introduced 
by means of D6-branes, and the backreaction generates a flow between two conformal theories in the UV and IR. The flow in that case can be determined to all orders in $N_f$, and was studied 
using various observables including the entanglement entropy of a disc. They extracted universal contributions using the renormalized entanglement proposal of Liu-Mezei. Even though the setup is sufficiently different that we cannot directly compare their results to ours, it is worth mentioning that a term that can be extracted analytically has the form 
$$S \sim  c_{UV} (m R)^{2b} $$
where $b$ is related to the dimension of the deformation operator (quark-antiquark bilinear), and 
$c_{UV} \sim N N_f / \sqrt{\lambda}$. So in that case also one has a coupling constant dependence in the coefficient. It would be worthwhile to check whether the same coefficient is obtained also for the term corresponding to the one we computed, which in a 3d theory is proportional to $m$. 

It would also be interesting to consider other cases of massive theories obtained by top-down string constructions (for instance in the flavored Klebanov-Strassler model \cite{Nunez:2010sf}), as well as considering the setup of $D3/D7$ branes at finite temperature and density;  the background geometries are known also in this case \cite{Bigazzi:2011it}. 

Finally, as we mentioned in the introduction, there are other mass-dependent terms with coefficients that depend on the geometry of the entangling region. In the case we studied we found one coefficient related to the curvature of the entangling surface, namely the constant term in (\ref{Sxi}), that is non zero for the ball. 
It would be interesting to compute the entanglement for other cases, e.g. in the case of a waveguide geometry. Unfortunately we have not been able to find an analytic solution for the corresponding equations of motion for the minimal surface.

\subsection*{Acknowlegments}
GP would like to thank E. Tonni for useful discussions, and the Isaac Newton Institute for Mathematical Sciences, Cambridge, for support and hospitality during the programme ``Mathematics and Physics of the Holographic Principle"where the final stages of this work were completed.  We would also like to thank 
 Y. Bea, E. Conde, N. Jokela and A. Ramallo for communications about the results of their paper.


\begin{thebibliography}{100}

\bibitem{Kitaev:2005dm}
  A.~Kitaev and J.~Preskill,
  ``Topological entanglement entropy,''
  Phys.\ Rev.\ Lett.\  {\bf 96} (2006) 110404
  [hep-th/0510092].

\bibitem{Calabrese:2004eu}
  P.~Calabrese and J.~L.~Cardy,
  ``Entanglement entropy and quantum field theory,''
  J.\ Stat.\ Mech.\  {\bf 0406} (2004) P06002
  [hep-th/0405152].

\bibitem{Casini:2009sr}
  H.~Casini and M.~Huerta,
  ``Entanglement entropy in free quantum field theory,''
  J.\ Phys.\ A {\bf 42} (2009) 504007
  [arXiv:0905.2562 [hep-th]].

\bibitem{Ryu:2006bv}
  S.~Ryu and T.~Takayanagi,
  ``Holographic derivation of entanglement entropy from AdS/CFT,''
  Phys.\ Rev.\ Lett.\  {\bf 96} (2006) 181602
  [hep-th/0603001].
  
\bibitem{Nishioka:2009un}
  T.~Nishioka, S.~Ryu and T.~Takayanagi,
  J.\ Phys.\ A {\bf 42} (2009) 504008
  [arXiv:0905.0932 [hep-th]].

\bibitem{Casini:2011kv}
  H.~Casini, M.~Huerta and R.~C.~Myers,
  ``Towards a derivation of holographic entanglement entropy,''
  JHEP {\bf 1105} (2011) 036
  [arXiv:1102.0440 [hep-th]].

\bibitem{Lewkowycz:2013nqa}
  A.~Lewkowycz and J.~Maldacena,
  ``Generalized gravitational entropy,''
  JHEP {\bf 1308} (2013) 090
  [arXiv:1304.4926 [hep-th]].

\bibitem{Solodukhin:2008dh}
  S.~N.~Solodukhin,
  ``Entanglement entropy, conformal invariance and extrinsic geometry,''
  Phys.\ Lett.\ B {\bf 665} (2008) 305
  [arXiv:0802.3117 [hep-th]].

\bibitem{Fursaev:2006ng}
  D.~V.~Fursaev,
  Phys.\ Rev.\ D {\bf 73} (2006) 124025
  [hep-th/0602134].

\bibitem{Hertzberg:2010uv}
  M.~P.~Hertzberg and F.~Wilczek,
  ``Some Calculable Contributions to Entanglement Entropy,''
  Phys.\ Rev.\ Lett.\  {\bf 106} (2011) 050404
  [arXiv:1007.0993 [hep-th]].

\bibitem{Hertzberg:2012mn}
  M.~P.~Hertzberg,
  ``Entanglement Entropy in Scalar Field Theory,''
  J.\ Phys.\ A {\bf 46} (2013) 015402
  [arXiv:1209.4646 [hep-th]].

\bibitem{Liu:2012eea}
  H.~Liu and M.~Mezei,
  ``A Refinement of entanglement entropy and the number of degrees of freedom,''
  JHEP {\bf 1304} (2013) 162
  [arXiv:1202.2070 [hep-th]].
  
\bibitem{Liu:2013una}
  H.~Liu and Már.~Mezei,
  ``Probing renormalization group flows using entanglement entropy,''
  arXiv:1309.6935 [hep-th].

\bibitem{Hung:2011ta}
  L.~-Y.~Hung, R.~C.~Myers and M.~Smolkin,
  ``Some Calculable Contributions to Holographic Entanglement Entropy,''
  JHEP {\bf 1108} (2011) 039
  [arXiv:1105.6055 [hep-th]].

\bibitem{Lewkowycz:2012qr}
  A.~Lewkowycz, R.~C.~Myers and M.~Smolkin,
  ``Observations on entanglement entropy in massive QFT's,''
  JHEP {\bf 1304} (2013) 017
  [arXiv:1210.6858 [hep-th]].

\bibitem{Bea:2013jxa}
  Y.~Bea, E.~Conde, N.~Jokela and A.~V.~Ramallo,
  ``Unquenched massive flavors and flows in Chern-Simons matter theories,''
  arXiv:1309.4453 [hep-th].

\bibitem{Karch:2003nh}
  A.~Karch and A.~Katz,
  ``Adding flavor to AdS/CFT,''
  Fortsch.\ Phys.\  {\bf 51} (2003) 759.

\bibitem{Chang:2013mca}
  H.~-C.~Chang and A.~Karch,
  ``Entanglement Entropy for Probe Branes,''
  arXiv:1307.5325 [hep-th].

\bibitem{Nunez:2010sf}
  C.~Nunez, A.~Paredes and A.~V.~Ramallo,
  ``Unquenched Flavor in the Gauge/Gravity Correspondence,''
  Adv.\ High Energy Phys.\  {\bf 2010} (2010) 196714
  [arXiv:1002.1088 [hep-th]].
  
\bibitem{Bigazzi:2009bk}
  F.~Bigazzi, A.~L.~Cotrone, J.~Mas, A.~Paredes, A.~V.~Ramallo and J.~Tarrio,
  ``D3-D7 Quark-Gluon Plasmas,''
  JHEP {\bf 0911} (2009) 117
  [arXiv:0909.2865 [hep-th]].

\bibitem{Magana:2012kh}
  A.~Magana, J.~Mas, L.~Mazzanti and J.~Tarrio,
  ``Probes on D3-D7 Quark-Gluon Plasmas,''
  JHEP {\bf 1207} (2012) 058
  [arXiv:1205.6176 [hep-th]].

\bibitem{Bigazzi:2011it}
  F.~Bigazzi, A.~L.~Cotrone, J.~Mas, D.~Mayerson and J.~Tarrio,
  ``D3-D7 Quark-Gluon Plasmas at Finite Baryon Density,''
  JHEP {\bf 1104} (2011) 060
  [arXiv:1101.3560 [hep-th]].
  



\end{thebibliography}
\end{document}